\input mn.tex
\input epsf.tex

%
%\Referee
%
\hyphenation{Pij-pers}
\hyphenation{mode-set}
%     
% EQUATION NUMBERING:
\newcount\eqnumber
\eqnumber=1
%\new macro produces sequentially numbered equations by writing \eqno\neqn)
%at end of displayed equations
\def\neqn{{\rm(\the\eqnumber)}\global\advance\eqnumber by 1}
%to refer to an equation which is 5 equations back, write "equation \ref5)"
\def\refeq#1){\advance\eqnumber by -#1 {\rm(\the\eqnumber)} \advance
\eqnumber by #1}
%to name an equation, place command "\eqnam{\Poisson}{Poisson}" before
%equation, and
%thereafter "equation (\Poisson)" will generate the proper equation number.
\def\eqnam#1#2{\immediate\write1{\xdef\ #2{(\the\eqnumber}}\xdef#1{(\the\eqnumber}}
% FIGURE NUMBERING:
\newcount\fignumber
\fignumber=1
%\nfig macro assigns number to a figure
\def\nfig{\global\advance\fignumber by 1}
\def\refig#1{\advance\fignumber by -#1 \the\fignumber \advance\fignumber by #1}
\def\fignam#1#2{\immediate\write1{\xdef\ #2{\the\fignumber}}\xdef#1{\the\fignumber}}
\def\note #1]{{\bf #1]}}
\def\draft{\headline{\bf File: \jobname\hfill DRAFT\hfill\today}}
\def\ref{\par\noindent
	\hangindent=0.7 truecm
	\hangafter=1}
%
%
% Redefine et al. , e.g., i.e., for A&A style
%
\def\spose#1{\hbox to 0pt{#1\hss}}
\def\pomega{\mathrel{\spose{\raise 2pt\hbox{$\mathchar"218$}}
     \hbox{$\omega$}}}
\def\ddr#1{{\partial #1\over\partial r}}
\def\dtdr#1{{\partial^2 #1\over\partial r^2}}
\def\ddu#1{{\partial #1\over\partial u}}
\def\etal{{et al.}}

%  To generate a "referee version" (single column, double-line spacing)
%  activate the command by deleting the "%" in the following line:
%\Referee
\loadboldmathnames

\begintopmatter

\title{Helioseismic determination of the solar gravitational quadrupole
moment}

\author{ Frank~P.~Pijpers}
\affiliation{Theoretical Astrophysics Center, Institute for Physics and
Astronomy, Aarhus University, Ny Munkegade, DK-8000~Aarhus~C, Denmark}
\shortauthor{F.P. Pijpers }
\shorttitle{the solar gravitational quadrupole}
\acceptedline{}

\abstract{ 
One of the most well-known tests of General Relativity (GR) results from
combining measurements of the anomalous precession of the orbit of Mercury
with a determination of the gravitational quadrupole moment of the Sun
$J_2$. The latter can be done by inference from an integral relation
between $J_2$ and the solar internal rotation.
New observational data of high quality obtained from the Solar Heliospheric
Satellite (SoHO) and from the Global Oscillations Network Group (GONG),
allow the determination of the internal rotation velocity of the Sun
as a function of radius and latitude with unprecedented spatial resolution
and accuracy. As a consequence, a number of global properties of the Sun
can also be determined with much higher accuracy, notably the gravitational
quadrupole moment of the Sun. The anomalous precession of the orbit of Mercury is 
primarily due to GR effects but there are classical corrections the largest of
which is that due to $J_2$. It is shown here that the data are currently
consistent with the predictions of GR.
}


\maketitle

\section{Introduction}

For observations that are well resolved in space and in time the 
oscillations of the Sun can be decomposed into its pulsation 
eigenmodes, which are products of functions of radius and
of spherical harmonic functions. Each mode, and therefore each measured
oscillation frequency, is uniquely identified by three numbers~: the radial
order $n$, and the degree $l$ and the azimuthal order $m$ of the spherical 
harmonic. The solar rotation produces oscillation frequencies that are split 
into multiplets. The relationship between the mode frequencies and the 
rotation is~:
\eqnam\Drotrel{Drotrel}
$$
2\pi {\nu_{nlm} - \nu_{nl\,-m}\over 2 m}\ =\ \int\limits_0^1 
\int\limits_{-1}^1\, {\rm d} x\, 
{\rm d}\cos \theta\, K_{nlm}(x,\theta) \Omega(x,\theta) 
\eqno\neqn
$$
where $x = r/R_\odot$ is the fractional radius, $R_\odot$ is the radius 
of the Sun, and $\theta$ the colatitude. The 
$K_{nlm}$ are the mode kernels for rotation. The Global Oscillations 
Network Group (GONG) produces values for the splittings through their 
data-reduction pipeline which are then available for inversion of the
above integral relation. The Solar Heliospheric Satellite (SoHO) SOI/MDI 
instrument pipe-line generally produces Ritzwoller-Lavely a-coefficients 
(Ritzwoller \& Lavely, 1991), instead of individual splittings. The 
relation between these a-coefficients and the rotation rate is a linear 
integral equation very similar to \Drotrel) although with different 
kernels. Explicit expressions for the kernels for both cases can be found 
in e.g. Pijpers (1997).

Using these data it is possible to determine the internal rotation rate 
of the Sun using inverse techniques. Results of such inversions can be
found in e.g. Thompson \etal{} (1996) and Schou \etal{} (1997). Apart from
the resolved rotation rate, there are some global properties of the Sun of
astrophysical interest, which are related to the internal rotation rate 
through integral equations. One of these quantities is the total angular 
momentum $H$ of the Sun which is related to the internal rotation rate 
through~:
\eqnam\AngMom{AngMom}
$$
H\equiv\int\limits_{0}^{1} {\rm d}x \int\limits_{-1}^{1} {\rm d}\cos\theta
\ {\cal I} \Omega (x, \theta)
\eqno\neqn$$
with the moment of inertia kernel ${\cal I}$~:
\eqnam\MomIn{MomIn}
$$
{\cal I} = 2\pi R_\odot^5 \rho x^4 (1 -\cos^2 \theta)
\eqno\neqn$$
where $\rho$ is the density inside the Sun.
Another is the total kinetic energy $T$ in rotation which is given by~:
\eqnam\KinEn{KinEn}
$$
T\equiv\int\limits_{0}^{1} {\rm d}x \int\limits_{-1}^{1} {\rm d}\cos\theta
\ {1\over 2}{\cal I} \Omega^2 (x, \theta)
\eqno\neqn$$
Since the total angular momentum is related linearly to the rotation rate
$\Omega$ it is possible to construct the kernel ${\cal I}$ directly from
a linear combination of the individual model kernels (or a-coefficient kernels)
using for instance the technique of Subtractive Optimally Localized Averages
(SOLA) (cf. Pijpers and Thompson, 1992, 1994) as was done using GONG data 
by Pijpers (1998). This avoids the circuitous route of first determining 
the resolved rotation rate and then re-integrating. The reasons for doing 
this are that it can have better properties from the point of view 
propagation of the measurement errors, as well as avoiding systematic
errors introduced at each computational step and it is computationally 
much less expensive. It is not possible to do the same for
the kinetic energy since this is quadratic in the rotation rate. To 
determine $T$, one has to either follow the route of determining the 
resolved rotation rate, taking the square and then re-integrating, or 
one must make use of the second order splittings which are generally 
much less well-determined and are affected by physical effects other 
than rotation such as internal magnetic fields (cf. Gough \& Thompson, 1990).

Another quantity of particular interest is the gravitational quadrupole 
moment $J_2$ of the Sun, caused by its flattening due to the rotation. 
The gravitational quadrupole moment $J_2$ of the Sun is that component
of the gravitational field corresponding to the second Legendre polynomial
as a function of co-latitude in an expansion of the gravitational field 
on Legendre polynomials. It is related to the solar oblateness 
$\Delta_\phi$, the ellipticity of the visible solar disk, as $J_2 = {2\over 3}
\Delta_\phi$.
The gravitational quadrupole moment of the Sun modifies the precession 
of the orbits of the planets. Therefore in using for instance the 
precession of the orbit of Mercury for testing the prediction from GR it is 
necessary to know $J_2$. Expressions for the integral relation between
$J_2$ and the internal rotation rate of the Sun have been derived for
special cases of a rotation rate dependent on the radius only, or on 
simple parameterizations with respect to latitude (cf. Gough, 1981, 1982 ;
Ulrich \& Hawkins, 1981). More general expressions have been given by
Dziembowski \& Goode (1992) who expand the rotation rate by projection 
onto Legendre polynomials. However it can be shown that this is a somewhat 
cumbersome approach and quite simple expressions can be found even for a
general distribution of $\Omega (x, \theta)$.

In section 2, the integral relation between the gravitational quadrupole 
moment of the Sun and a general internal rotation rate is given. In section 3
the results are given of performing the direct inversion for $H$ and 
the values for $T$ and $J_2$ obtained by taking the square of the resolved
rotation rate and re-integrating, using data from GONG and using data from
SOI/MDI on board SoHO. Conclusions are presented in section 4.

\section{The gravitational quadrupole moment}

The general expressions relating the various moments of the gravitational 
potential of rotating stars to their rotation rate have been given by
Goldreich \& Schubert (1968) and by Lebovitz (1970). These lead to what is
essentially Clairaut-Legendre equations for the moments. For 
convenience the steps will be briefly repeated here.
Starting point are Poisson's equation which relates the gravitational 
potential to the density distribution~:
\eqnam\Poisson{Poisson}
$$
\nabla^2 \phi = -4 \pi G \rho\ ,
\eqno\neqn$$
and the equation of motion~:
\eqnam\motion{motion}
$$
\rho\nabla\phi = \nabla p - \rho\Omega (r, \theta)^2 r \sin\theta {\bf\pomega}
\eqno\neqn$$
where $\phi$ is the gravitational potential, $G$ is the constant of gravity,
$\rho$ and $p$ are the gas density and pressure respectively, and
$\Omega$ is the rotation rate which is a function of radius $r$ and
co-latitude $\theta$. ${\bf\pomega}$ is a unit vector perpendicular to
the rotation axis.
Writing equation \motion) out in components yields~:
\eqnam\eqmomcomp{eqmomcomp}
$$
\eqalign{
\rho \ddr{\phi} &= \ddr{p} - \rho r (1- u^2) \Omega (r,u)^2 \cr
\rho \ddu{\phi} &= \ddu{p} + \rho r^2 u \Omega (r,u)^2 \cr
}
\eqno\neqn$$
in which $u\equiv\cos\theta$.
Following the treatment of Goldreich \& Schubert (1968) and Lebovitz (1970)
for slowly rotating stars all quantities are described in terms of
perturbations of the spherically symmetric non-rotating star, i.e. 
$\Omega^2$ is treated as a quantity of first order in a small parameter 
expansion.
Subscripts $0$ refer to the non-rotating configuration, and $1$ to the
perturbed quantities. Collecting the first order terms in the
perturbation analysis of equation \eqmomcomp)~:
\eqnam\eqmomfo{eqmomfo}
$$
\eqalign{
\rho_0\ddr{\phi_1} + \rho_1\ddr{\phi_0} &= \ddr{p_1} - \rho_0 r (1- u^2)
\Omega (r,u)^2 \cr
\rho_0\ddu{\phi_1} &= \ddu{p_1} + \rho_0 r^2 u \Omega (r,u)^2 \cr
}\eqno\neqn$$
Of interest for the quadrupole moment of the gravitational potential is
the projection onto the Legendre polynomial $P_2 (u) = (3u^2-1)/2$.
In the first equation of \eqmomfo) all terms are multiplied by
${5\over 2} P_2 (u)$ and then integrated over $u$. The second equation would
yield $0=0$ since all its terms are odd in $u$. Therefore this equation
is first integrated in $u$ and then projected.
In the following the subscripts $12$ refer to the part of the first order 
perturbed quantities corresponding to these $P_2$ Legendre polynomial 
projections.
\eqnam\eqmomfop{eqmomfop}
$$
\eqalign{
\rho_0\ddr{\phi_{12}} + \rho_{12}\ddr{\phi_0} &= \ddr{p_{12}} -
\rho_0 r \int\limits_{-1}^{1}{\rm d}u\ {5\over 3} \left[1-P_2(u)\right]
\cr
&\hskip 1cm\times P_2(u)\Omega(r,u)^2 \cr
\rho_0 \phi_{12} - p_{12} &= \rho_0 r^2 \int\limits_{-1}^{1}{\rm d}u\ {5\over 2}
P_2(u) \int\limits_{-1}^{u}{\rm d}v\, v\Omega(r,v)^2
}\eqno\neqn$$
The double integral in the second equation can be re-written, using partial
integration~:
$$
\eqalign{
\int\limits_{-1}^{1}{\rm d}u\ {5\over 2} P_2(u) &\int\limits_{-1}^{u}{\rm d}v
\ v\Omega(r,v)^2 = \cr 
&= {5\over 4}\left\{\left[ \left( u^3 - u \right) \int\limits_{-1}^{u}{\rm d}v
\ v\Omega(r,v)^2 \right]_{-1}^{1} \right.\cr 
&\hskip 1cm \left. - \int\limits_{-1}^1{\rm d}u
\left( u^3 - u \right) u\Omega(r,u)^2\right\} \cr
&= {5\over 4} \int\limits_{-1}^1{\rm d}u \left( u^2 - u^4 \right)
\Omega(r,u)^2 \cr
}\eqno\neqn$$
The second equality of equation \eqmomfop) can be used to eliminate
$p_{12}$ from the first of \eqmomfop).
After some rearranging the result is~:
\eqnam\elimrho{elimrho}
$$
\eqalign{
\rho_{12}\ddr{\phi_0} &= \phi_{12} \ddr{\rho_0} - \ddr{}\left[ \rho_0 r^2
{\cal G}(\Omega)\right] +\cr 
&\hskip 12pt -\rho_0 r \int\limits_{-1}^1{\rm d} u\ {5\over 3}
\left(1-P_2(u)\right)P_2(u)\Omega(r,u)^2 \cr
p_{12} &= \rho_0\phi_{12} - \rho_0 r^2 {\cal G}(\Omega) \cr
}\eqno\neqn$$
where ${\cal G}$ is defined by~:
\eqnam\Gdef{Gdef}
$$
{\cal G}(\Omega) \equiv {5\over 4} \int\limits_{-1}^{1}{\rm d}u\
\left( u^2 - u^4 \right) \Omega(r,u)^2
\eqno\neqn$$
The relevant equations from the perturbed Poisson's equation \Poisson) are~:
$$
\eqalign{
{1\over r^2}\ddr{}\left(r^2\ddr{\phi_0}\right) &= -4\pi G\rho_0 \cr
\dtdr{\phi_{12}} + {2\over r}\ddr{\phi_{12}}- {6\over r^2}\phi_{12} &=
-4 \pi G \rho_{12} \cr
}\eqno\neqn$$
in which $\rho_{12}$ can now be substituted using the first of equations
\elimrho)~:
\eqnam\inhde{inhde}
$$
\eqalign{
\dtdr{\phi_{12}} + &{2\over r}\ddr{\phi_{12}}- {6\over r^2}\phi_{12}
= {4\pi r^2\over {\cal M}_r}\biggl\{ \phi_{12}\ddr{\rho_0} \cr
& -\ddr{}\left[\rho_0 r^2 {\cal G}(\Omega)\right] \cr
&\left. - \rho_0 r \int\limits_{-1}^1{\rm d} u\ {5\over 3}
\left(1-P_2(u)\right)P_2(u)\Omega(r,u)^2 \right\}\cr }
\eqno\neqn$$
in which use has been made of the mass within radius $r$~:
$$
{\cal M}_r \equiv \int\limits_0^r{\rm d}r'\ 4\pi\rho_0 r'^2
\eqno\neqn$$
Now define the linear differential operator ${\cal L}$~:
$$
{\cal L}\phi_{12}\ \equiv\ \left\{\ddr{}\left(r^2\ddr{}\right) -
\left( 6 + {4\pi r^4\over{\cal M}_r}\ddr{\rho_0}\right)\right\}
\phi_{12}
\eqno\neqn$$
and a function $f$~:
$$
\eqalign{
f(r) &\equiv -{4\pi r^4\over {\cal M}_r}\left\{ r^2\ddr{}\left[
\rho_0{\cal G}(\Omega)\right] + \rho_0 r \biggl[ 2{\cal G}(\Omega)+
\right.\cr 
&\hskip 1cm \left.\left.
\int\limits_{-1}^1{\rm d} u\ {5\over 3}
\left(1-P_2(u)\right)P_2(u)\Omega(r,u)^2 \right]\right\} \cr
&= -{4\pi r^4\over {\cal M}_r}\left\{ r^2\ddr{}\left[
\rho_0{\cal G}(\Omega)\right] - \rho_0 r \times \right.\cr
&\hskip 1cm\left.\int\limits_{-1}^1{\rm d} u\
{5\over 4} \left( u^2 -1 \right)\left(5u^2 -1\right) \Omega(r,u)^2
\right\} \cr
}\eqno\neqn$$
so that $\phi_{12}$ is the solution of ${\cal L}\phi_{12} = f(r).$
This equation can be solved used Green's functions. For $r>R_{\odot}$
the density $\rho_0 \equiv 0$ and therefore $f(r) \equiv 0$. An exact
solution is then $\phi_{12} = r^{-3}$.
If another solution $\psi$ of ${\cal L}\psi = 0$ is constructed which
is regular at $r=0$ the general solution is~:
\eqnam\intrel{intrel}
$$
\phi_{12} (R) = \int\limits_0^{\infty}{\rm d}z\  G(R,z) f(z)
\eqno\neqn$$
with the Green's function~:
$$
G(R,z) = \cases{ \displaystyle {\psi(R) z^{-3} \over z^2 W(z)}
\hskip 2cm 0\leq R \leq z\cr
\displaystyle {\psi(z) R^{-3} \over z^2 W(z)} \hskip 2cm 0\leq z \leq R\cr}
\eqno\neqn$$
where $W(z)$ is the Wronskian of the solutions $r^{-3}$ and $\psi$~:
$$
W(z) = \left|\matrix{ \psi & z^{-3} \cr \psi' & -3 z^{-4} \cr}\right|
 = -z^{-6}{{\rm d}\over {\rm d} z} (z^3 \psi)
\eqno\neqn$$
Since $r^{-3}$ is not a solution of ${\cal L}\psi = 0$ for $r<R_{\odot}$ this
equation is {\bf only} valid for $R\geq R_{\odot}$. Of interest here 
is the solution $\phi_{12}$ at $R=R_{\odot}$. If in
\intrel) $R$ is replaced with $R_{\odot}$ it is allowed to replace $z^2 W(z)$
with $R_{\odot}^2 W(R_{\odot})$, which can be verified by substitution
of \intrel) into \inhde).
Since $f(r) = 0$ for $r>R_{\odot}$, the expression for $\phi_{12}(R_\odot)$ 
simplifies to~:
$$
\phi_{12} (R_\odot) = -R_{\odot}^{-3} \left[ r^{-4}{{\rm d}\over {\rm d} r}
(r^3 \psi)\right]^{-1}_{r=R_{\odot}} \int\limits_{0}^{R_{\odot}}{\rm d}z\
\psi (z) f(z)
\eqno\neqn$$
The solar oblateness $\Delta_{\phi}$ is related to $\phi_{12}$ as
$$
\Delta_{\phi} = -{3\over 2} {R_{\odot} \over G {\cal M}_{\odot}}
\phi_{12} (R_{\odot})
\eqno\neqn$$
Substituting the expressions for $\phi_{12}$ and $f(r)$~:
\eqnam\quadru{quadru}
$$
\eqalign{
\Delta_{\phi} = &{2\pi R_{\odot}^2 \over G {\cal M}_{\odot}} \left[
{{\rm d}\over {\rm d} r}(r^3 \psi) \right]^{-1}_{r=R_{\odot}}
\int\limits_0^{R_\odot}{\rm d}r\ \left\{{-3r^6 \psi(r) \over {\cal M}_r}
\times\right.\cr
&\ddr{}\left[\rho_0 {\cal G}\right] + 
{3 r^6 \psi(r) \over {\cal M}_r} {\rho_0\over r} \left[
\int\limits_{-1}^1{\rm d} u\ {5\over 4} \left( u^2 -1 \right)\times\right.\cr
&\hskip 1cm\left.\left(5u^2 -1\right) \Omega(r,u)^2 \right] \biggr\} \cr
=&{2\pi R_{\odot}^2 \over G {\cal M}_{\odot}} \left[
{{\rm d}\over {\rm d} r}(r^3 \psi) \right]^{-1}_{r=R_{\odot}}
\int\limits_0^{R_\odot}{\rm d}r\ \biggl\{ 3\rho_0 {\cal G} \times\cr
&\ddr{}\left({r^6\psi(r)\over {\cal M}_r}\right) + 
{3 r^6 \psi(r) \over {\cal M}_r} {\rho_0\over r} \times\cr 
&\left.\left[
\int\limits_{-1}^1{\rm d} u\ {5\over 4} \left( u^2 -1 \right)\left(5u^2
-1\right) \Omega(r,u)^2 \right] \right\} \cr
= &{2\pi R_{\odot}^2 \over G {\cal M}_{\odot}} \left[
{{\rm d}\over {\rm d} r}(r^3 \psi) \right]^{-1}_{r=R_{\odot}}
\int\limits_0^{R_\odot}{\rm d}r\int\limits_{-1}^1{\rm d} u\ \cr
&\hskip 5mm\left[
{15\over 4}\rho_0 \ddr{}\left({r^6\psi(r)\over {\cal M}_r}\right) (u^2 -u^4)
+ \right.\cr
&\hskip 1cm\left.{15\over 4}{r^6 \psi(r) \over {\cal M}_r}
{\rho_0\over r} \left(u^2 -1 \right)\left( 5u^2-1\right)\right]\Omega(r,u)^2 \cr
}\eqno\neqn$$
in which the second equality is obtained by partial integration, and the
third is a re-arranging of terms making use of the definition \Gdef) of
${\cal G}$. 

For an $\Omega$ which is a function of $r$ only, integration over $u$
of the second term between square brackets in \quadru) is identical to
$0$ and the first term reduces to $\rho_0 \ddr{}\left({r^6\psi(r)\over
{\cal M}_r}\right) \Omega(r,u)^2$.
Equation \quadru) then reduces to equation (12) of Gough (1981).
Once the density $\rho_0(r)$ of the Sun is known, it is
trivial to calculate the two-dimensional kernel~:
$$
\eqalign{
{\cal F} (r, u) \equiv &{15\pi R_{\odot}^2 \over 2 G {\cal M}_{\odot}} \left[
{{\rm d}\over {\rm d} r}(r^3 \psi) \right]^{-1}_{r=R_{\odot}}
{\rho_0 \over r}\left({r^6\psi(r)\over {\cal M}_r}\right) \times\cr
&\hskip 6mm\left[{\partial \ln \left({r^6\psi(r)\over {\cal M}_r}\right)\over
\partial\ln r} u^2 - \left( 5u^2-1\right)\right]\left( 1-u^2\right)
\cr}
\eqno\neqn$$
Determining $\Delta_\phi$ is thus reduced to evaluating the two-dimensional
integral~:
$$
\Delta_\phi = \int\limits_0^{R_\odot}{\rm d}r\int\limits_{-1}^1{\rm d} u\
{\cal F} (r, u) \Omega(r,u)^2
\eqno\neqn$$
Direct inversion would have to make use of the second order splittings 
in an inverse problem, so the same route is followed as with the kinetic 
energy $T$~: the resolved $\Omega^2$ is multiplied with ${\cal F}$ and
integrated.

\fignam{\kernel}{kernel}
\beginfigure{1} 
%\vskip -11.0cm\hskip 6mm
\epsfysize=11.0cm
\epsfbox{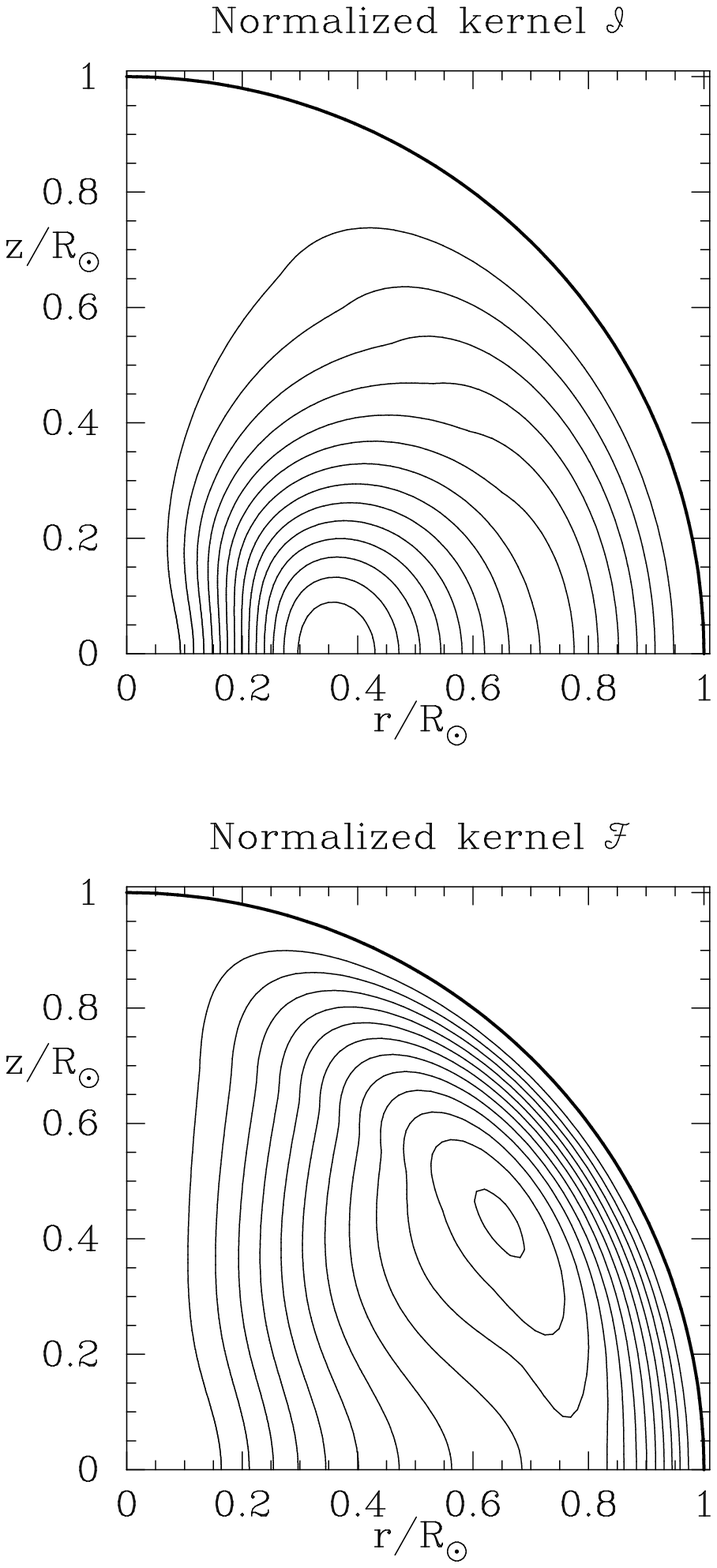}
\caption{{\bf Figure \kernel.}
{The normalized kernel ${\cal I}$ for determination of 
the total solar angular momentum and kinetic energy (top panel), and 
the kernel ${\cal F}$ for the determination of the gravitational 
quadrupole moment of the Sun. Contour intervals are 0.1 for both kernels.
}}
\endfigure\nfig

Using the standard solar model S of Christensen-Dalsgaard (cf. 
Christensen-Dalsgaard \etal{}, 1996) the kernels
${\cal I}$ and ${\cal F}$ were calculated, and normalized to have unit 
integral over the solar volume. Contour plots in one quadrant are shown
in figure \kernel{}. The other quadrants can be obtained by reflection 
in the coordinate axes. 

\section{Results and conclusions}

Two independent data sets have been used to determine the total solar 
angular momentum, the total kinetic energy and the gravitational 
quadrupole moment.
One dataset is in the form of splittings obtained with the earth-based 
GONG network of telescopes~: 33169 splittings distributed over 542 
complete multiplets with $7 \leq l \leq 150$ and $1.5\ {\rm mHz} < 
\nu < 3.5\ {\rm mHz}$ gathered from GONG months 4 to 10. The other 
dataset is in the form of a-coefficients gathered from 144 d out of the 
first 6 months of 
operation of the SOI/MDI instrument on board the SoHO satellite. The 
data consists of 414 multiplets with $1 \leq l \leq 250$ and $1.0\ {\rm mHz} < 
\nu < 4.2\ {\rm mHz}$, and the odd a-coefficients up to at most $a_{35}$ 
are available.

The GONG data leads to the values~:
$$
\eqalign{
H_{\rm d} &= \left[186.3 \pm 2.4 \right] 10^{39}\ {\rm kg\ m^2\ s^{-1}} \cr
H_{\rm i} &= \left[186.3 \pm 3.7 \right] 10^{39}\ {\rm kg\ m^2\ s^{-1}} \cr
T &= \left[245.5 \pm 9.8 \right] 10^{33}\ {\rm kg\ m^2\ s^{-2}} \cr
J_2 &= \left[ 2.14 \pm 0.09 \right] 10^{-7} \cr
}\eqno\neqn$$
The MDI data leads to the values~:
$$
\eqalign{
H_{\rm d} &= \left[192.3 \pm 1.9 \right] 10^{39}\ {\rm kg\ m^2\ s^{-1}} \cr
H_{\rm i} &= \left[192.9 \pm 3.9 \right] 10^{39}\ {\rm kg\ m^2\ s^{-1}} \cr
T &= \left[262.5 \pm 10. \right] 10^{33}\ {\rm kg\ m^2\ s^{-2}} \cr
J_2 &= \left[2.23 \pm 0.09 \right] 10^{-7} \cr
}\eqno\neqn$$
The subscript d refers to a determination directly from the data using the
freedom of the SOLA method to construct the kernel directly, the subscript
i refers to the indirect method, which is re-integrating the resolved
$\Omega$. $T$ and $J_2$ have been determined by re-integration only.
Since the direct method should suffer much less from systematic effects,
the value $H_{\rm i}$ is shown merely to demonstrate consistency between the 
two methods. Error weighted means for $H_{\rm d}, T,$ and $J_2$ are~:
\eqnam\ewmHTJ{ewmHTJ}
$$
\eqalign{
H &= \left[ 190.0 \pm 1.5 \right] 10^{39}\ {\rm kg\ m^2\ s^{-1}} \cr
T &= \left[ 253.4 \pm 7.2 \right] 10^{33}\ {\rm kg\ m^2\ s^{-2}} \cr
J_2 &= \left[2.18 \pm 0.06 \right] 10^{-7} \cr
}\eqno\neqn$$
This determination of $J_2$ is entirely consistent with that of Patern\'o
\etal{} (1996) who used direct oblateness measurements of the solar disk
to infer the quadrupole moment.

One of the most well-known tests of GR results from
combining measurements of the precession of the orbit of Mercury
(cf. Shapiro \etal{}, 1976~; Anderson \etal{}, 1987, 1991, 1992)
with a determination of the gravitational quadrupole moment of the Sun
$J_2$. In the fully conservative parameterized post-newtonian (PPN) 
formalism, the predicted advance $\Delta\phi_0$ per orbital period of a 
planetary orbit with semi-major axis $a$ and eccentricity $e$, after 
correcting for perturbations due to other planets, is~:
\eqnam\precpp{precpp}
$$
\Delta\phi_0 = {6\pi G M \lambda_p \over a(1-e^2)c^2}
\eqno\neqn$$
where
\eqnam\lambeq{lambeq}
$$
\lambda_p = {1\over 3} (2-\beta+2\gamma)+ {R^2 c^2 \over 2 G M a (1-e^2)} J_2
\eqno\neqn$$
Here $M$ and $R$ are the mass and radius of the Sun, $G$ is the gravitational
constant, and $c$ is the speed of light. The parameters $\beta$ and 
$\gamma$ are the Eddington-Robertson parameters of the PPN formalism 
(cf. Misner \etal, 1973), which in general relativity are equal to $1$.
For Mercury the above relation \precpp) reduces to~:
$$
\Delta\phi_0 = 42.9794\ \lambda_p\ "/{\rm century}
\eqno\neqn$$
and \lambeq) is~:
$$
\lambda_p = {1\over 3} (2-\beta+2\gamma)+ 2.96\ 10^3 \times J_2
\eqno\neqn
$$
Shapiro \etal{} (1976), using planetary radar ranging, found an anomalous
precession for Mercury's orbit of $43.11 \pm 0.21$. Using radar and 
spacecraft ranging Anderson \etal{} (1987) found $42.92 \pm 0.20$ and 
an update (Anderson \etal{}, 1991) gives the value $42.94 \pm 0.20$. 
The most recent result reported by Anderson \etal{} (1992) is $43.13 \pm 0.14$. 
Combining the most recent value for the anomalous precession of Mercury's 
orbit and the value for $J_2$ given above in equations \ewmHTJ) yields~:
\eqnam\lambval{lambval}
$$
{1\over 3} (2-\beta+2\gamma) = 1.003 \pm 0.003 
\eqno\neqn$$
The error quoted here is entirely due to that in the planetary ranging 
data, since the error due to the uncertainty in $J_2$ is two orders of 
magnitude smaller.

In this paper it is thus demonstrated that the total solar angular momentum,
its total kinetic energy in rotation, and the solar gravitational 
quadrupole moment can be determined through inverting integral equations
that are linear in the rotation rate $\Omega$ or in its square, with known
integration kernels. 
The value of the gravitational quadrupole moment \ewmHTJ) when combined with 
planetary ranging data for the precession of the orbit of Mercury yields
a value for the combined PPN formalism parameters \lambval) 
which is consistent with GR in which this combination is 
predicted to be exactly equal to unity. More stringent tests of GR using 
the orbit of Mercury rely on measuring its orbital precession with much 
greater precision.

\section*{Acknowledgments}
The Theoretical Astrophysics Center is a collaboration 
between Copenhagen University and Aarhus University and is funded by 
Danmarks Grundforskningsfonden. GONG is managed by NSO, a division of 
NOAO that is operated by the Association of Universities for Research 
in Astronomy under co-operative agreement with NSF. The GONG data were 
acquired by instruments operated by the BBSO, HAO, Learmonth, Udaipur, 
IAC, and CTIO. The MDI project operating the SOI/MDI experiment on board 
the SoHO spacecraft is supported by NASA contract NAG5-3077 at Stanford 
University.

\section*{References}
\ref
Anderson J.D., Campbell J.K., Jurgens R.F., Lau E.L., Newhall X X,
Slade III M.A., Standish Jr. E.M., 1992,
Recent Developments in Solar-System Tests of General Relativity, 
Proceedings of The Sixth Marcel Grossmann Meeting,
Eds. H. Sato and T. Nakamura, World Scientific Publishing Co., Singapore
\ref
Anderson J.D., Colombo G., Espsitio P.B., Lau E.L., Trager G.B., 1987,
{Icarus}
71, 337
\ref
Anderson J.D., Slade M.A., Jurgens R.F., Lau E.L., Newhall X X,
Standish Jr. E.M., 1991,
{Pub. Astron. Soc. Australia}
9, 324
\ref
Christensen-Dalsgaard J., \etal{}, 1996,
{Science}
272, 1286
\ref
Dziembowski W.A., Goode P.R., 1992,
{ApJ}
394, 670
\ref
Goldreich P., Schubert G., 1968,
{ApJ}
154, 1005
\ref
Gough D.O., 1981,
{MNRAS}
196, 731
\ref
Gough D.O., 1982,
{Nature}
298, 334
\ref
Gough D.O, Thompson M., 1990,
{MNRAS}
242, 25
\ref
Lebovitz N.R., 1970,
{ApJ}
160, 701
\ref
Misner C., Thorne K.S., Wheeler J.A., 1973,
`Gravitation',
Freeman, San Francisco, 1072, 1116
\ref
Patern\'o L., Sofia S., DiMauro M.P., 1996,
{A\&A}
314, 940
\ref
Pijpers F.P., 1997,
{A\&A}
326, 1235
\ref
\ref
Pijpers F.P., 1998,
{Proc. IAU symposium 181, 'Sounding Solar and Stellar Interiors'}
Eds. J. Provost, F.-X. Schmider, Kluwer, Dordrecht, in press
\ref
Pijpers F.P., Thompson M.J., 1992,
{A\&A}
262, L33
\ref
Pijpers F.P., Thompson M.J., 1994,
{A\&A}
281, 231
\ref
Pijpers F.P., Thompson M.J., 1996,
{MNRAS}
279, 498
\ref
Ritzwoller M.H., Lavely E.M., 1991,
{ApJ}
369, 557
\ref
Schou J., \etal{}, 1997,
{ApJ}
submitted
\ref
Shapiro I.I., Counselman C.C. III, King R.W., 1976,
Phys. Rev. Lett.,
36, 555
\ref
Thompson M., \etal{}, 1996,
{Science}
272, 1300
\ref
Ulrich R.K, Hawkins G.W., 1981,
{ApJ},
246, 985
%\ref
%Schou J., Christensen-Dalsgaard J., Thompson M.J., 1994,
%{ApJ}
%433, 389 (SCDT)

\bye